\documentclass[11pt]{article}
\usepackage[margin=1.1in]{geometry}
\usepackage{booktabs}
\usepackage{amsmath}
\usepackage{graphicx}
\usepackage[hidelinks]{hyperref}
\usepackage{microtype}

\newcommand{\AbRaUbcfFiveFull}{+0.0107 [+0.0033, +0.0189]}

\newcommand{\AbTlUbcfFiveFull}{\ensuremath{-}0.0096 [\ensuremath{-}0.0162, \ensuremath{-}0.0036]}

\newcommand{\ApFloor}{0.1643}

\newcommand{\ApPop}{0.3129}

\newcommand{\InHighlights}{300}
\newcommand{\InPerStratum}{50}
\newcommand{\InRelAgree}{147}
\newcommand{\InRelDiffer}{3}
\newcommand{\InUnits}{150}

\newcommand{\RaNarrowApKnnOneFull}{+0.0887 [+0.0795, +0.0982]}

\newcommand{\RaNarrowApUbcfOneFull}{+0.0922 [+0.0817, +0.1026]}

\newcommand{\RsPosTasteApKnnOneFull}{\ensuremath{-}0.0099 [\ensuremath{-}0.0155, \ensuremath{-}0.0040]}

\newcommand{\RsPosTasteApUbcfOneFull}{\ensuremath{-}0.0153 [\ensuremath{-}0.0209, \ensuremath{-}0.0095]}

\newcommand{\RsTasteModApKnnOneFull}{\ensuremath{-}0.0162 [\ensuremath{-}0.0191, \ensuremath{-}0.0134]}

\newcommand{\RsTasteModApUbcfOneFull}{\ensuremath{-}0.0192 [\ensuremath{-}0.0223, \ensuremath{-}0.0162]}

\newcommand{\RtAnchoredOneShare}{0.765}
\newcommand{\RtAnchoredRunShare}{0.168}
\newcommand{\RtAnchoredScatterShare}{0.067}
\newcommand{\RtAnchoredTotal}{193{,}904}

\newcommand{\RtApKnnVsProxFortyFiveFull}{\ensuremath{-}0.0234 [\ensuremath{-}0.0330, \ensuremath{-}0.0144]}

\newcommand{\RtApKnnVsProxFortyOneFull}{\ensuremath{-}0.0157 [\ensuremath{-}0.0221, \ensuremath{-}0.0097]}

\newcommand{\RtApKnnVsProxFortyThreeFull}{\ensuremath{-}0.0134 [\ensuremath{-}0.0207, \ensuremath{-}0.0064]}

\newcommand{\RtApProxBlendFortyVsProxFortyFiveFull}{+0.0008 [\ensuremath{-}0.0015, +0.0032]}

\newcommand{\RtApProxBlendFortyVsProxFortyOneFull}{+0.0024 [+0.0005, +0.0043]}

\newcommand{\RtApProxBlendFortyVsProxFortyThreeFull}{+0.0024 [+0.0004, +0.0045]}

\newcommand{\RtApProxFortyVsProxTwentyOneFull}{+0.0076 [+0.0044, +0.0109]}

\newcommand{\RtApUbcfVsProxFortyFiveFull}{\ensuremath{-}0.0214 [\ensuremath{-}0.0307, \ensuremath{-}0.0130]}

\newcommand{\RtApUbcfVsProxFortyOneFull}{\ensuremath{-}0.0144 [\ensuremath{-}0.0205, \ensuremath{-}0.0084]}

\newcommand{\RtApUbcfVsProxFortyThreeFull}{\ensuremath{-}0.0124 [\ensuremath{-}0.0193, \ensuremath{-}0.0051]}

\newcommand{\RtHitNextKnnVsProxFortyFiveFull}{\ensuremath{-}0.1539 [\ensuremath{-}0.1828, \ensuremath{-}0.1254]}

\newcommand{\RtHitNextKnnVsProxFortyOneFull}{\ensuremath{-}0.1248 [\ensuremath{-}0.1432, \ensuremath{-}0.1072]}

\newcommand{\RtHitNextKnnVsProxFortyThreeFull}{\ensuremath{-}0.1247 [\ensuremath{-}0.1464, \ensuremath{-}0.1039]}

\newcommand{\RtHitNextProxBlendFortyVsProxFortyFiveFull}{\ensuremath{-}0.0127 [\ensuremath{-}0.0236, \ensuremath{-}0.0026]}

\newcommand{\RtHitNextProxBlendFortyVsProxFortyOneFull}{\ensuremath{-}0.0017 [\ensuremath{-}0.0092, +0.0057]}

\newcommand{\RtHitNextProxBlendFortyVsProxFortyThreeFull}{\ensuremath{-}0.0064 [\ensuremath{-}0.0153, +0.0022]}

\newcommand{\RtHitNextUbcfVsProxFortyFiveFull}{\ensuremath{-}0.1396 [\ensuremath{-}0.1689, \ensuremath{-}0.1121]}

\newcommand{\RtHitNextUbcfVsProxFortyOneFull}{\ensuremath{-}0.1152 [\ensuremath{-}0.1339, \ensuremath{-}0.0975]}

\newcommand{\RtHitNextUbcfVsProxFortyThreeFull}{\ensuremath{-}0.1129 [\ensuremath{-}0.1345, \ensuremath{-}0.0904]}

\newcommand{\RtLateApProxFortyOneFull}{+0.0128 [\ensuremath{-}0.0101, +0.0378]}

\newcommand{\RtLateApUbcfVsProxFortyOneFull}{+0.0054 [\ensuremath{-}0.0219, +0.0314]}

\newcommand{\RtLateHitNextAheadAbsOne}{0.5070}

\newcommand{\RtLateHitNextAheadOneFull}{+0.2773 [+0.1579, +0.3953]}

\newcommand{\RtLateHitNextPopAbsOne}{0.2297}

\newcommand{\RtLateHitNextUbcfVsAheadOneFull}{\ensuremath{-}0.1771 [\ensuremath{-}0.2941, \ensuremath{-}0.0584]}

\newcommand{\RtLateNOne}{365}

\newcommand{\RtRestHitNextAheadOne}{\ensuremath{-}0.02}

\newcommand{\RtRestNOne}{1{,}027}

\newcommand{\RtSubContigHitNextAheadOneFull}{+0.2533 [+0.2219, +0.2862]}

\newcommand{\RtSubContigHitNextUbcfVsAheadOneFull}{\ensuremath{-}0.2223 [\ensuremath{-}0.2546, \ensuremath{-}0.1900]}

\newcommand{\RtSubContigNOne}{6{,}316}

\newcommand{\RtsNarrowApProxBlendFortyVsProxFortyOneFull}{+0.0136 [+0.0107, +0.0165]}

\newcommand{\RtsPosTasteApProxBlendFortyVsProxFortyOneFull}{\ensuremath{-}0.0025 [\ensuremath{-}0.0040, \ensuremath{-}0.0011]}

\newcommand{\RtsTasteModApProxBlendFortyVsProxFortyOneFull}{+0.0041 [+0.0030, +0.0052]}

\newcommand{\RvHitKnnFiveFull}{+0.0106 [\ensuremath{-}0.0077, +0.0292]}

\newcommand{\RvHitKnnOneFull}{+0.0201 [+0.0075, +0.0325]}

\newcommand{\RvHitKnnThreeFull}{+0.0218 [+0.0070, +0.0368]}

\newcommand{\RvHitUbcfFiveFull}{+0.0249 [+0.0074, +0.0428]}

\newcommand{\RvHitUbcfOneFull}{+0.0297 [+0.0150, +0.0449]}

\newcommand{\RvHitUbcfThreeFull}{+0.0336 [+0.0172, +0.0494]}

\newcommand{\RvHlMinusTlKnnFiveFull}{+0.0135 [+0.0073, +0.0200]}

\newcommand{\RvHlMinusTlUbcfFiveFull}{+0.0167 [+0.0106, +0.0232]}

\newcommand{\RvHlMinusTlUbcfThreeFull}{+0.0094 [+0.0040, +0.0144]}

\newcommand{\RwApAheadFiveFull}{+0.0107 [\ensuremath{-}0.0086, +0.0294]}

\newcommand{\RwApAheadOneFull}{\ensuremath{-}0.0269 [\ensuremath{-}0.0413, \ensuremath{-}0.0116]}

\newcommand{\RwApAheadThreeFull}{\ensuremath{-}0.0014 [\ensuremath{-}0.0174, +0.0153]}

\newcommand{\RwApPopFive}{0.3188}
\newcommand{\RwApPopOne}{0.3129}

\newcommand{\RwApPopThree}{0.3154}

\newcommand{\RwApProxFortyFiveFull}{+0.0154 [+0.0073, +0.0240]}

\newcommand{\RwApProxFortyOneFull}{+0.0126 [+0.0066, +0.0187]}

\newcommand{\RwApProxFortyThreeFull}{+0.0085 [+0.0018, +0.0162]}

\newcommand{\RwApProxTenFiveFull}{+0.0023 [\ensuremath{-}0.0095, +0.0148]}

\newcommand{\RwApProxTenOneFull}{\ensuremath{-}0.0069 [\ensuremath{-}0.0169, +0.0040]}

\newcommand{\RwApProxTenThreeFull}{\ensuremath{-}0.0059 [\ensuremath{-}0.0165, +0.0054]}

\newcommand{\RwApProxTwentyFiveFull}{+0.0112 [+0.0014, +0.0217]}

\newcommand{\RwApProxTwentyOneFull}{+0.0050 [\ensuremath{-}0.0030, +0.0136]}

\newcommand{\RwApProxTwentyThreeFull}{+0.0032 [\ensuremath{-}0.0065, +0.0121]}

\newcommand{\RwApUbcfVsProxTwentyFiveFull}{\ensuremath{-}0.0172 [\ensuremath{-}0.0285, \ensuremath{-}0.0078]}

\newcommand{\RwApUbcfVsProxTwentyOneFull}{\ensuremath{-}0.0068 [\ensuremath{-}0.0143, \ensuremath{-}0.0004]}

\newcommand{\RwBelowOne}{0.851}

\newcommand{\RwCandOne}{96}

\newcommand{\RwFixN}{2{,}743}

\newcommand{\RwFixOneKnnApLastVsFirstFull}{+0.0084 [+0.0015, +0.0152]}

\newcommand{\RwFixOneProxApLastVsFirstFull}{+0.0765 [+0.0645, +0.0889]}

\newcommand{\RwFixOneUbcfApLastVsFirstFull}{+0.0129 [+0.0048, +0.0212]}

\newcommand{\RwFunNoPosOne}{572}

\newcommand{\RwFunReadersOne}{20{,}450}

\newcommand{\RwFunShortOne}{4{,}122}

\newcommand{\RwFunSmallPoolOne}{5{,}709}

\newcommand{\RwFunUnanchOne}{2{,}704}

\newcommand{\RwHitAllAheadOne}{\ensuremath{-}0.0922}

\newcommand{\RwHitAllAheadOneFull}{\ensuremath{-}0.0922 [\ensuremath{-}0.1208, \ensuremath{-}0.0632]}

\newcommand{\RwHitNextAheadAbsFive}{0.5654}
\newcommand{\RwHitNextAheadAbsOne}{0.4712}
\newcommand{\RwHitNextAheadAbsThree}{0.5304}

\newcommand{\RwHitNextAheadFiveFull}{+0.3522 [+0.3110, +0.3911]}

\newcommand{\RwHitNextAheadOne}{+0.2147}

\newcommand{\RwHitNextAheadOneFull}{+0.2147 [+0.1836, +0.2458]}

\newcommand{\RwHitNextAheadPctOne}{47\%}

\newcommand{\RwHitNextAheadThreeFull}{+0.3015 [+0.2639, +0.3388]}

\newcommand{\RwHitNextPopFive}{0.2132}
\newcommand{\RwHitNextPopOne}{0.2565}
\newcommand{\RwHitNextPopPctOne}{26\%}

\newcommand{\RwHitNextPopThree}{0.2289}

\newcommand{\RwHitNextProxFortyFiveFull}{+0.1645 [+0.1402, +0.1913]}

\newcommand{\RwHitNextProxFortyOneFull}{+0.1449 [+0.1276, +0.1619]}

\newcommand{\RwHitNextProxFortyThreeFull}{+0.1465 [+0.1262, +0.1673]}

\newcommand{\RwHitNextProxTenFiveFull}{+0.2690 [+0.2343, +0.3028]}

\newcommand{\RwHitNextProxTenOneFull}{+0.2037 [+0.1804, +0.2253]}

\newcommand{\RwHitNextProxTenThreeFull}{+0.2271 [+0.2010, +0.2546]}

\newcommand{\RwHitNextProxTwentyFiveFull}{+0.2180 [+0.1872, +0.2466]}

\newcommand{\RwHitNextProxTwentyOneFull}{+0.1829 [+0.1628, +0.2036]}

\newcommand{\RwHitNextProxTwentyThreeFull}{+0.1932 [+0.1696, +0.2162]}

\newcommand{\RwHitNextUbcfPctOne}{29\%}

\newcommand{\RwHitNextUbcfVsAheadFiveFull}{\ensuremath{-}0.3273 [\ensuremath{-}0.3687, \ensuremath{-}0.2828]}

\newcommand{\RwHitNextUbcfVsAheadOneFull}{\ensuremath{-}0.1850 [\ensuremath{-}0.2174, \ensuremath{-}0.1539]}

\newcommand{\RwHitNextUbcfVsAheadThreeFull}{\ensuremath{-}0.2680 [\ensuremath{-}0.3066, \ensuremath{-}0.2300]}

\newcommand{\RwHitNextUbcfVsProxTwentyOneFull}{\ensuremath{-}0.1532 [\ensuremath{-}0.1739, \ensuremath{-}0.1323]}

\newcommand{\RwNextSentOne}{1.7}

\newcommand{\RwNextShareKnnOne}{9\%}

\newcommand{\RwNextShareUbcfOne}{14\%}

\newcommand{\RwRandHitOne}{0.0919}

\newcommand{\RwSimManyOne}{0.510}

\newcommand{\RwUnitsFive}{2{,}743}
\newcommand{\RwUnitsOne}{7{,}343}
\newcommand{\RwUnitsThree}{4{,}240}

\newcommand{\RwWithinThreeOne}{0.410}

\newcommand{\RxFuzzyShare}{0.060}

\newcommand{\TmAnchored}{213{,}656}
\newcommand{\TmAnchoredPages}{1{,}741}
\newcommand{\TmAnchoredShare}{0.775}

\newcommand{\TmGroups}{2{,}434}
\newcommand{\TmGroupsAll}{4{,}661}
\newcommand{\TmGroupsIdentical}{46}
\newcommand{\TmGroupsNoText}{2{,}181}
\newcommand{\TmHighlights}{275{,}814}

\newcommand{\TmKnnFiveFull}{\ensuremath{-}0.0080 [\ensuremath{-}0.0154, \ensuremath{-}0.0009]}

\newcommand{\TmKnnFiveHolm}{0.2860}

\newcommand{\TmKnnFourFull}{\ensuremath{-}0.0061 [\ensuremath{-}0.0129, +0.0006]}

\newcommand{\TmKnnFourHolm}{0.7120}

\newcommand{\TmKnnOneFull}{\ensuremath{-}0.0031 [\ensuremath{-}0.0079, +0.0019]}

\newcommand{\TmKnnOneHolm}{1.0000}

\newcommand{\TmKnnRandOneFull}{+0.0108 [+0.0066, +0.0150]}

\newcommand{\TmKnnRegpopOneFull}{\ensuremath{-}0.0070 [\ensuremath{-}0.0116, \ensuremath{-}0.0025]}

\newcommand{\TmKnnThreeFull}{\ensuremath{-}0.0049 [\ensuremath{-}0.0114, +0.0013]}

\newcommand{\TmKnnThreeHolm}{0.8932}

\newcommand{\TmKnnTwoFull}{\ensuremath{-}0.0048 [\ensuremath{-}0.0101, +0.0005]}

\newcommand{\TmKnnTwoHolm}{0.7120}

\newcommand{\TmMaxUpper}{+0.0037}
\newcommand{\TmNoTextShare}{0.468}

\newcommand{\TmPOneFull}{\ensuremath{-}0.0024 [\ensuremath{-}0.0133, +0.0085]}

\newcommand{\TmPages}{1{,}511}

\newcommand{\TmUbcfFiveFull}{\ensuremath{-}0.0060 [\ensuremath{-}0.0126, +0.0006]}

\newcommand{\TmUbcfFiveHolm}{0.7120}

\newcommand{\TmUbcfFourFull}{\ensuremath{-}0.0040 [\ensuremath{-}0.0108, +0.0028]}

\newcommand{\TmUbcfFourHolm}{1.0000}

\newcommand{\TmUbcfOneFull}{\ensuremath{-}0.0018 [\ensuremath{-}0.0072, +0.0037]}

\newcommand{\TmUbcfOneHolm}{1.0000}

\newcommand{\TmUbcfRegpopOneFull}{\ensuremath{-}0.0057 [\ensuremath{-}0.0104, \ensuremath{-}0.0011]}

\newcommand{\TmUbcfThreeFull}{\ensuremath{-}0.0039 [\ensuremath{-}0.0103, +0.0025]}

\newcommand{\TmUbcfThreeHolm}{1.0000}

\newcommand{\TmUbcfTwoFull}{\ensuremath{-}0.0042 [\ensuremath{-}0.0097, +0.0013]}

\newcommand{\TmUbcfTwoHolm}{0.8932}

\newcommand{\TmUnitsFive}{2{,}743}
\newcommand{\TmUnitsFour}{3{,}406}
\newcommand{\TmUnitsOne}{7{,}343}
\newcommand{\TmUnitsThree}{4{,}240}
\newcommand{\TmUnitsTwo}{5{,}443}

\title{Reading Position Is the Baseline to Beat:\\A Time-Ordered Evaluation of Personalised Highlight Prediction}
\author{Kazuki Nakayashiki \qquad Keisuke Watanabe\\[2pt]
Glasp Inc.\\
\texttt{\{kazuki,kei\}@glasp.co}}
\date{}

\begin{document}
\maketitle

\begin{abstract}
A reader's first highlights on a page are the cheapest personal signal a reading product has. The
natural plan is to suggest what similar earlier readers marked, and to judge the result against
popularity. We argue that the baseline to beat is reading position. In a time-ordered evaluation on
one social highlighting platform (\TmUnitsOne{} reader-page pairs on \TmPages{} pages after one
highlight), ranking the sentences just below a reader's first highlight, with no other reader's
data, puts the next highlight in the top five \RwHitNextAheadPctOne{} of the time, against
\RwHitNextPopPctOne{} for popularity and \RwHitNextUbcfPctOne{} for the better of two similarity
methods. The baseline depends on the target: over all later highlights that ranking loses to
popularity, while popularity discounted by distance from the latest highlight, at the scale with
the best average precision of three tried, beats popularity and both similarity methods on both
targets. In a comparison
specified in advance, neither similarity method shows a gain over popularity in average precision
over all later highlights, from one to five highlights, and a gain of +0.01 is excluded. Nor would a
gain by itself show that a method has found a reader's preferences: synthetic readers who share one
set of preferences produce one, and an evaluation out of time order shows a method where the reader
went. The position results are exploratory and unconfirmed. Personalisation inside a document should
be evaluated in time order and against reading position.
\end{abstract}

\section{Introduction}

Social highlighting platforms let readers see which passages other readers marked. The crowd's most
marked passages are an obvious thing to suggest to a reader who has just arrived on a page. A more
ambitious suggestion is personal: once the reader has made a highlight or two, find the earlier
readers who marked the same things and suggest what \emph{they} marked. This is user-based
collaborative filtering under the ``given-$m$'' protocol~\cite{breese1998}, applied inside one
document, and the natural test of it is one comparison: does it beat popularity?

That comparison leaves out the simplest thing a reader's first highlights reveal. They carry two
kinds of information about the reader. One is \emph{position}: reading is sequential, a highlight is
made where the reader is, and the next one is usually made a little further down the page. The other
is \emph{preference}: what this reader likes to mark, which is what a method that looks for similar
earlier readers hopes to find. Both are personal, and adapting to either is personalisation. But they
are different things, and a comparison with popularity cannot tell them apart. An earlier reader who
shares a mark with the newcomer was reading the same part of the page, so a method that looks for
similar readers finds, first of all, readers who were at the same place, and its gain over popularity
can come from position and not from preference. (We do not observe where a reader is reading. We
observe where they last highlighted, and use that as a proxy for their position throughout.)

\textbf{This paper makes one claim: inside a document, reading position is the baseline that
personalisation from a reader's first highlights has to beat.} We support it with a time-ordered
evaluation on one platform, in which a method sees only what existed when the reader made their
$m$-th highlight. That readers move down a page is not news. The contribution is what happens to the
evaluation of personalisation once position is in it.
\begin{enumerate}
\item \textbf{The baseline changes the verdict.} Take a reader's next highlight. Against
  popularity, the better of two standard similarity methods looks like a modest success: after one
  highlight it puts the next one in its top five \RwHitNextUbcfPctOne{} of the time, against
  \RwHitNextPopPctOne{} for popularity. Against position it is far behind: a rule that ranks the
  sentences just below that highlight, with no other reader's data, does so \RwHitNextAheadPctOne{}
  of the time (\S\ref{sec:position}). Over all later highlights the two methods do not beat
  popularity even modestly: in a comparison specified in advance, neither shows a gain in average
  precision from one to five highlights, and a gain of +0.01 is excluded (\S\ref{sec:main}).
\item \textbf{The target decides which baseline.} After one highlight, the rule that is far ahead for
  the next highlight is behind popularity over everything the reader highlights later: on the same
  units and the same metric (Hit@5), its difference from popularity is \RwHitNextAheadOne{} for the
  next highlight and \RwHitAllAheadOne{} over all later ones. There the baseline to beat is
  popularity discounted by distance from the reader's position: at the scale with the best average
  precision of the three we tried, both similarity methods are below it on both targets
  (\S\ref{sec:position}). The baseline to beat cannot be named before the prediction target is.
\item \textbf{A gain does not by itself show preference.} Blended into the position baseline,
  similarity adds about two thousandths of average precision. On synthetic pages such a gain appears
  where all readers share one set of preferences, and can be absent where they do not
  (\S\ref{sec:taste}). An evaluation out of time order compounds the problem, because it shows a
  method where the reader went (\S\ref{sec:protocol}).
\end{enumerate}

The comparison with popularity was specified in advance; the position results are exploratory and
have not been confirmed on new records. Every result is conditional on the readers a time-ordered
evaluation can score: those whose $m$-th highlight comes after at least five other readers have
highlighted the page, and whose first highlights can be matched to the page's text as it is today.
And nothing here says that readers lack distinct preferences: our contrasts cannot separate
preference from position, in either direction.

\section{Related work}

\paragraph{Personalising highlights.} On the same platform, re-ranking the sentences of a document
with a personal model built from a reader's history on other documents did not beat an impersonal
salience order~\cite{selection}, and a second study found that readers of one document form
sub-groups and proposed the test we run~\cite{factions, beyond26}. With far more input than ours, a
matrix-factorisation model given a reader's highlights on two passages of a textbook predicted their
highlights on a third better than the other readers' highlighting rate~\cite{winchell2020}; that
study ignored the order of highlighting and compared with popularity alone.

\paragraph{Strong simple baselines.} Simple baselines are often hard to beat: neighbourhood methods
remain competitive in session-based recommendation~\cite{jannach2017, ludewig2018}, and of seven
reproducible neural recommenders six could often be outperformed by comparably simple
methods~\cite{ferrari2019}. Our point is about which baseline the comparison needs: inside a
document, popularity is not the only simple one.

\paragraph{Position as a baseline.} Where items come in an order, position alone is a strong
predictor, and it can be mistaken for something else. In web search, users click results ranked
higher more often, even when those results are less relevant than others they looked
at~\cite{joachims2005}. In news summarisation, the first three sentences of an article are a
baseline that a neural summariser did not surpass on ROUGE~\cite{see2017} and that later systems
outperformed only slightly~\cite{kryscinski2019}, and it has been questioned whether such models
learn to identify important content or only to find the start of the document~\cite{kedzie2018}. In
an online course, predicting that a learner goes to the next page in the course's own order was
right about half the time~\cite{pardos2017}. On our platform the crowd's highlights are
front-loaded, so that truncating a document is a strong predictor of them~\cite{fusion}. The
position we study is of another kind: not where an item is displayed or where a sentence stands in a
document, but where one reader has got to, as revealed by that reader's latest highlight. To our
knowledge, personalised highlight prediction has not been evaluated against it.

\paragraph{Evaluation in time order.} Offline protocols that ignore the global timeline let a model
use interactions that would not exist at prediction time~\cite{ji2023}, and the choice of data split
changes which models win~\cite{meng2020, gusak2025}. We find the same hazard inside a single
document, where what may leak is the reader's own later position (\S\ref{sec:protocol}).

\section{Setup}\label{sec:setup}

\paragraph{Task.} A page is a sequence of sentences, and its readers highlight passages of it over
time; the sentences a reader has highlighted are that reader's \emph{marks}. Fix a reader who is new
to the page, though not necessarily to the service (a \emph{newcomer}), and a number $m$. At the
time of the newcomer's $m$-th highlight, the \emph{cut}, a scorer is given the sentences of the
newcomer's first $m$ highlights (the \emph{reveal}) and every other reader's highlights made before
the cut. It ranks the sentences not yet revealed (the \emph{candidates}). Nothing made after the
cut, by anyone, reaches the scorer. We score two targets: \emph{all later highlights}, the sentences
the newcomer highlights after the cut (the registered target), and the \emph{next highlight}, the
newcomer's first later highlight that adds an unrevealed sentence.

\paragraph{Data.} Readers mark passages on web pages; marks are public by default and a reader can
see other readers' marks. The data are a read-only extraction of 2026-10-04 from one platform: the
pages that pass the platform's own rules for public article pages and have at least six readers with
a public highlight document, with the URL variants of a page merged. Each reader's highlights are
kept in the order of their recorded times;
imported, suggested and image highlights are dropped. A page's text is fetched now and split into
sentences, and a highlight is \emph{anchored} to every sentence that contains it, that it contains,
or whose token Jaccard with it is at least 0.6. The evaluation therefore keeps the order of events,
but it does not reconstruct the page as each reader saw it.

A unit is a (page, newcomer, $m$). It needs the newcomer's first $m$ highlights all to anchor, at
least one later highlight that adds a sentence, and a \emph{pool} of at least five other readers
with a highlight before the cut. That leaves \TmUnitsOne{} units on \TmPages{} pages at $m=1$
(Table~\ref{tab:family} has the other $m$, Appendix~\ref{app:data} the filters and the counts).

One of the authors checked a sample of the records by hand: \InUnits{} units, comparing the text of
each reader's first highlight and of their next highlight with the sentences the two were anchored
to. All \InHighlights{} highlights were anchored where their text is; the two highlights were
ordered as their recorded times in every unit; pages merged from several URLs showed one article in
every case; and the next highlight lay below, above or across the first, as its anchors say, in
\InRelAgree{} of the \InUnits{} units (Appendix~\ref{app:data}); in the other \InRelDiffer{} the two
readings differ only on whether it overlaps the first or lies wholly below it.

\paragraph{Scorers.} Table~\ref{tab:scorers} lists them. A newcomer with no positively similar pool
reader gets the POP ranking from KNN and UBCF. The position scorers take the newcomer's
\emph{position} to be the sentences of their latest revealed highlight: the distance $d$ of a
sentence is the number of sentences to the nearest of them, and ``below'' means after the last of
them. AHEAD looks at no other reader; it ranks the sentences below the position first and goes on
with those above. PROXPOP($\tau$) uses every pool reader and chooses none; we tried three scales,
$\tau=10$, 20 and 40 sentences, and named 20 as the default before computing.

\begin{table}[t]
\centering\footnotesize\setlength{\tabcolsep}{5pt}
\begin{tabular}{@{}l p{0.60\linewidth} p{0.17\linewidth}@{}}
\toprule
scorer & ranks the unrevealed sentences by & pool readers used \\
\midrule
\multicolumn{3}{@{}l}{\emph{Popularity}} \\
POP & the number of pool readers who marked the sentence & all \\
\multicolumn{3}{@{}l}{\emph{Reader similarity}} \\
KNN & the marks of the $k=3$ pool readers most similar to the reveal (positive similarity only) & up to three \\
UBCF & every pool reader's marks, weighted by similarity to the reveal & all, weighted \\
\multicolumn{3}{@{}l}{\emph{Reading position}} \\
AHEAD & the sentences below the newcomer's position first, nearest first & none \\
PROXPOP($\tau$) & $\mathrm{POP}\cdot e^{-d/\tau}$, with $d$ the distance from the newcomer's position & all \\
\bottomrule
\end{tabular}
\caption{The scorers. Similarity is the binary cosine between the revealed sentences and a pool
reader's marks. The newcomer's position is their latest revealed highlight, and distances are in
sentences. POP, KNN and UBCF were registered; the position scorers were added in exploratory
analyses.}
\label{tab:scorers}
\end{table}

\paragraph{Metrics and inference.} The registered metric is average precision (AP) over the
candidates, with all later highlights as positives. For the next highlight we report Hit@5, the
probability that the top five hold at least one of its sentences. Both are exact expectations under
a uniformly random order within tied scores~\cite{mcsherry2008}. Scorers are compared by paired
differences on the same units. Intervals are 95\% percentile bootstraps
that resample pages and readers independently with replacement (a two-way, or pigeonhole,
bootstrap~\cite{owen2007}), because the same readers appear on many pages, and two-sided $p$-values
are twice the smaller share of resamples on either side of zero. We call an estimate \emph{resolved}
when its interval excludes zero.

\paragraph{Registered and exploratory.} The comparison of the similarity scorers with popularity
(\S\ref{sec:main}) is \emph{registered}: its specification was committed to the authors' private
repository, and its code and inputs hash-frozen, before any of its outcomes was computed. It is a
family of eleven tests, Holm-adjusted, with 10,000 resamples, and it fixed $+0.01$ AP in advance as
the smallest gain we would act on. Everything in \S\ref{sec:position}--\ref{sec:protocol} is
exploratory: computed after the registered result was known, on the same units (the
\emph{registered units}) unless a subset is named, with 2,000 resamples and no adjustment for the
number of intervals. We therefore put most weight on patterns that hold across $m$.

\section{Over all later highlights, neither similarity scorer beats popularity}\label{sec:main}

Table~\ref{tab:family} gives the registered comparison. Every estimate is negative, no Holm-adjusted
$p$ is below 0.05, and no upper bound exceeds \TmMaxUpper{}, so a gain of +0.01 is excluded at
$m=1$--5. Mean AP at $m=1$ is \ApPop{} for POP, against a random floor of \ApFloor{}. The family's
eleventh test contrasts KNN $-$ POP at $m=1$ between the top and
the bottom third of an agreement statistic of the earlier readers~\cite{factions}; it is not resolved
either (\TmPOneFull{}).

\begin{table}[t]
\centering\small\setlength{\tabcolsep}{5pt}
\begin{tabular}{rrlrlr}
\toprule
$m$ & units & KNN $-$ POP & Holm $p$ & UBCF $-$ POP & Holm $p$ \\
\midrule
1 & \TmUnitsOne & \TmKnnOneFull & \TmKnnOneHolm & \TmUbcfOneFull & \TmUbcfOneHolm \\
2 & \TmUnitsTwo & \TmKnnTwoFull & \TmKnnTwoHolm & \TmUbcfTwoFull & \TmUbcfTwoHolm \\
3 & \TmUnitsThree & \TmKnnThreeFull & \TmKnnThreeHolm & \TmUbcfThreeFull & \TmUbcfThreeHolm \\
4 & \TmUnitsFour & \TmKnnFourFull & \TmKnnFourHolm & \TmUbcfFourFull & \TmUbcfFourHolm \\
5 & \TmUnitsFive & \TmKnnFiveFull & \TmKnnFiveHolm & \TmUbcfFiveFull & \TmUbcfFiveHolm \\
\bottomrule
\end{tabular}
\caption{The registered comparison: average precision over all later highlights, in time order
(estimate and 95\% two-way interval; Holm over the family's eleven tests, 10,000 resamples). The
eleventh test is in the text.}
\label{tab:family}
\end{table}

The first highlight is not uninformative. In a registered secondary contrast (unadjusted), KNN beats
as many random pool readers by \TmKnnRandOneFull{} at $m=1$: the highlight does pick out readers who
predict better than random ones. On this target they show no gain over the crowd. Exploratory checks
find no detectable gain elsewhere either: on the units where three or more pool readers share a mark
with the reveal (\RwSimManyOne{} of units at $m=1$), on the units and pages where anchoring is best,
and with other neighbour choices, denser pools or more highlights revealed (ancillary files), though
several of these subsets are too thin to exclude a gain of +0.01.

The registered analysis also carried one position-aware baseline: popularity counted among the pool
readers who marked in the 20-sentence blocks of the newcomer's revealed highlights. In registered
secondary contrasts (unadjusted) both similarity scorers are below it at $m=1$ and 5 (at $m=1$ the
differences are \TmUbcfRegpopOneFull{} for UBCF and \TmKnnRegpopOneFull{} for KNN).

\section{Reading position as the baseline: alone for the next highlight, with popularity for all later ones}\label{sec:position}

\begin{figure}[t]
\centering
\includegraphics[width=\linewidth]{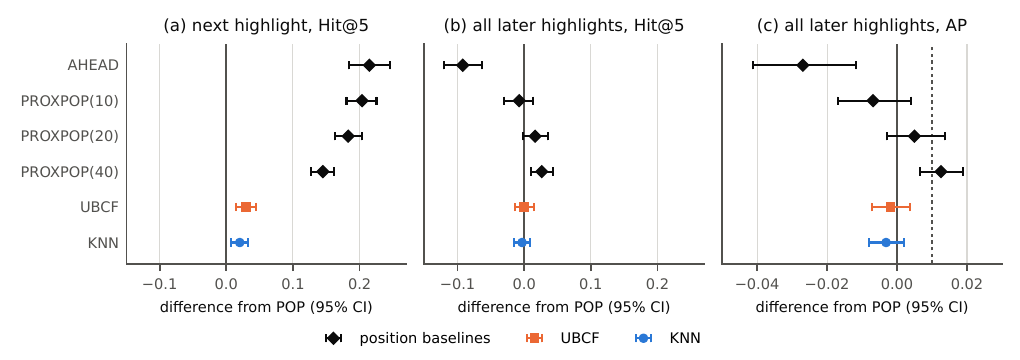}
\caption{The same scorers on two targets, at $m=1$: each scorer's difference from POP on the
registered units (95\% two-way intervals). Diamonds are position baselines; the square and the
circle are the two similarity scorers. (a) and (b) use the same metric on the same scale: position
alone (AHEAD) is far ahead for the next highlight and behind over all later highlights, and position
with popularity is resolved ahead on both only at the widest scale, PROXPOP(40). (c) is the
registered metric, with the registered estimates of Table~\ref{tab:family} for KNN and UBCF and the
pre-set threshold of +0.01 AP as a dotted line. Everything else is exploratory and unadjusted.}
\label{fig:scorers}
\end{figure}

Figure~\ref{fig:scorers} and Table~\ref{tab:position} put the newcomer's own position into the
comparison, without and with the crowd.

\subsection{The next highlight}

A unit at $m=1$ has a median of \RwCandOne{} candidate sentences, and the next highlight adds
\RwNextSentOne{} of them on average, so a random order would put it in the top five with probability
\RwRandHitOne{}. Popularity does so with probability \RwHitNextPopOne{}. AHEAD, which looks at no
other reader, does so with probability \RwHitNextAheadAbsOne{}, \RwHitNextAheadAbsThree{} and
\RwHitNextAheadAbsFive{} at $m=1$, 3 and 5: a gain over popularity of \RwHitNextAheadOneFull{} at
$m=1$. Readers move down the page: the sentences the next highlight adds all lie below the latest
highlight in \RwBelowOne{} of units, and the nearest of them is within three sentences of it in
\RwWithinThreeOne{}.

\begin{table}[t]
\centering\footnotesize\setlength{\tabcolsep}{4pt}
\begin{tabular}{llll}
\toprule
 & $m=1$ & $m=3$ & $m=5$ \\
\midrule
\multicolumn{4}{l}{\emph{Hit@5 for the next highlight: difference from POP}} \\
POP (mean) & \RwHitNextPopOne & \RwHitNextPopThree & \RwHitNextPopFive \\
AHEAD & \RwHitNextAheadOneFull & \RwHitNextAheadThreeFull & \RwHitNextAheadFiveFull \\
PROXPOP(10) & \RwHitNextProxTenOneFull & \RwHitNextProxTenThreeFull & \RwHitNextProxTenFiveFull \\
PROXPOP(20) & \RwHitNextProxTwentyOneFull & \RwHitNextProxTwentyThreeFull & \RwHitNextProxTwentyFiveFull \\
PROXPOP(40) & \RwHitNextProxFortyOneFull & \RwHitNextProxFortyThreeFull & \RwHitNextProxFortyFiveFull \\
UBCF & \RvHitUbcfOneFull & \RvHitUbcfThreeFull & \RvHitUbcfFiveFull \\
KNN & \RvHitKnnOneFull & \RvHitKnnThreeFull & \RvHitKnnFiveFull \\
\midrule
\multicolumn{4}{l}{\emph{AP over all later highlights: difference from POP}} \\
POP (mean) & \RwApPopOne & \RwApPopThree & \RwApPopFive \\
AHEAD & \RwApAheadOneFull & \RwApAheadThreeFull & \RwApAheadFiveFull \\
PROXPOP(10) & \RwApProxTenOneFull & \RwApProxTenThreeFull & \RwApProxTenFiveFull \\
PROXPOP(20) & \RwApProxTwentyOneFull & \RwApProxTwentyThreeFull & \RwApProxTwentyFiveFull \\
PROXPOP(40) & \RwApProxFortyOneFull & \RwApProxFortyThreeFull & \RwApProxFortyFiveFull \\
UBCF & \TmUbcfOneFull & \TmUbcfThreeFull & \TmUbcfFiveFull \\
KNN & \TmKnnOneFull & \TmKnnThreeFull & \TmKnnFiveFull \\
\bottomrule
\end{tabular}
\caption{Position and similarity against popularity on the registered units, in time order
(\RwUnitsOne{}, \RwUnitsThree{} and \RwUnitsFive{} units at $m=1$, 3 and 5). KNN and UBCF on AP are
the registered estimates of Table~\ref{tab:family}; everything else is exploratory (two-way
intervals, 2,000 resamples, unadjusted).}
\label{tab:position}
\end{table}

The similarity scorers do gain on this target, and far less. UBCF gains \RvHitUbcfOneFull{} over
popularity at $m=1$ and KNN \RvHitKnnOneFull{}, which is \RwNextShareUbcfOne{} and
\RwNextShareKnnOne{} of AHEAD's gain, and UBCF $-$ AHEAD is \RwHitNextUbcfVsAheadOneFull{},
\RwHitNextUbcfVsAheadThreeFull{} and \RwHitNextUbcfVsAheadFiveFull{} at $m=1$, 3 and 5. A method that
searches the earlier readers for the newcomer's likes is beaten, by a wide margin, by a rule that
knows only where the newcomer is.

\subsection{Everything the reader highlights later}

Over all later highlights, position alone is not enough: AHEAD is below popularity on AP at $m=1$
(\RwApAheadOneFull{}) and not detectably different at $m=3$ and 5. So the same rule is far ahead on
one target and behind on the other, on the same metric and the same units: on Hit@5 at $m=1$,
AHEAD $-$ POP is \RwHitNextAheadOneFull{} for the next highlight and \RwHitAllAheadOneFull{} over all
later highlights (Figure~\ref{fig:scorers}a and b).

Position \emph{with} popularity beats popularity, and both similarity scorers. PROXPOP(40),
popularity discounted by distance from the newcomer's position, beats POP on AP by
\RwApProxFortyOneFull{}, \RwApProxFortyThreeFull{} and \RwApProxFortyFiveFull{} at $m=1$, 3 and 5,
and it chooses no reader. Both similarity scorers are below it at each of these $m$, on both targets
(Table~\ref{tab:similarity}): at $m=1$, UBCF $-$ PROXPOP(40) is \RtApUbcfVsProxFortyOneFull{} on AP
and \RtHitNextUbcfVsProxFortyOneFull{} on Hit@5 for the next highlight.

\begin{table}[t]
\centering\footnotesize\setlength{\tabcolsep}{4pt}
\begin{tabular}{llll}
\toprule
 & $m=1$ & $m=3$ & $m=5$ \\
\midrule
\multicolumn{4}{l}{\emph{Scorer minus PROXPOP(40): AP over all later highlights}} \\
UBCF & \RtApUbcfVsProxFortyOneFull & \RtApUbcfVsProxFortyThreeFull & \RtApUbcfVsProxFortyFiveFull \\
KNN & \RtApKnnVsProxFortyOneFull & \RtApKnnVsProxFortyThreeFull & \RtApKnnVsProxFortyFiveFull \\
PROXBLEND(40) & \RtApProxBlendFortyVsProxFortyOneFull & \RtApProxBlendFortyVsProxFortyThreeFull & \RtApProxBlendFortyVsProxFortyFiveFull \\
\midrule
\multicolumn{4}{l}{\emph{Scorer minus PROXPOP(40): Hit@5 for the next highlight}} \\
UBCF & \RtHitNextUbcfVsProxFortyOneFull & \RtHitNextUbcfVsProxFortyThreeFull & \RtHitNextUbcfVsProxFortyFiveFull \\
KNN & \RtHitNextKnnVsProxFortyOneFull & \RtHitNextKnnVsProxFortyThreeFull & \RtHitNextKnnVsProxFortyFiveFull \\
PROXBLEND(40) & \RtHitNextProxBlendFortyVsProxFortyOneFull & \RtHitNextProxBlendFortyVsProxFortyThreeFull & \RtHitNextProxBlendFortyVsProxFortyFiveFull \\
\bottomrule
\end{tabular}
\caption{Similarity against, and blended into, the position baseline: paired differences from
PROXPOP(40) on the registered units (two-way intervals, 2,000 resamples, unadjusted). PROXBLEND(40)
ranks by $0.8\,B/\max B+0.2\,\mathrm{UBCF}/\max\mathrm{UBCF}$ with $B=\mathrm{PROXPOP}(40)$, the
maxima taken over the page's sentences; the weight was chosen for a blend with plain popularity and
not tuned again.}
\label{tab:similarity}
\end{table}

The scale was picked after the results: $\tau=40$ did best on AP of the three we tried, so its gain
over popularity, small as it is, is an optimistic choice among three. With the scale we had named in
advance, PROXPOP(20), the gain over POP is resolved only at $m=5$. Against that scale both
similarity scorers are still below on AP at $m=1$ and 5 (UBCF $-$ PROXPOP(20)
\RwApUbcfVsProxTwentyOneFull{} and \RwApUbcfVsProxTwentyFiveFull{}; at $m=3$ the interval includes
zero) and for the next highlight at every $m$ (\RwHitNextUbcfVsProxTwentyOneFull{} at $m=1$). On
neither target is either similarity scorer detectably above PROXPOP(20) or PROXPOP(40) at $m=1$, 3
or 5.

\subsection{A later slice, and anchoring}

Two checks ask whether the next-highlight result depends on when the highlights were made or on how
they anchor. Neither is a confirmation: both use parts of the data the result was found on. On the
\RtLateNOne{} units at $m=1$ whose first highlight came after a split on 2026-08-03, the last two
months of the data, AHEAD's Hit@5 is \RtLateHitNextAheadAbsOne{} against \RtLateHitNextPopAbsOne{}
for popularity (AHEAD $-$ POP \RtLateHitNextAheadOneFull{}; UBCF $-$ AHEAD
\RtLateHitNextUbcfVsAheadOneFull{}). On AP the slice resolves neither contrast of the previous
subsection (PROXPOP(40) $-$ POP \RtLateApProxFortyOneFull{}; UBCF $-$ PROXPOP(40)
\RtLateApUbcfVsProxFortyOneFull{}).

The position scorers take the newcomer's position from an anchor. On the \RtSubContigNOne{} units at
$m=1$ whose latest revealed highlight and next highlight each anchor to one sentence or one
contiguous run, AHEAD $-$ POP on next-highlight Hit@5 is \RtSubContigHitNextAheadOneFull{} and
UBCF $-$ AHEAD \RtSubContigHitNextUbcfVsAheadOneFull{}. On the other \RtRestNOne{} units, where at
least one of the two anchors to sentences that are not contiguous, AHEAD $-$ POP is about
\RtRestHitNextAheadOne{} by subtraction; we computed no interval for it.

\section{A gain would not by itself show preference}\label{sec:taste}

\paragraph{On our pages similarity adds little.} Blended into the position baseline, similarity adds
about two thousandths of AP at $m=1$ and 3 (Table~\ref{tab:similarity}: PROXBLEND(40) $-$ PROXPOP(40)
is \RtApProxBlendFortyVsProxFortyOneFull{} and \RtApProxBlendFortyVsProxFortyThreeFull{}) and nothing
detectable at $m=5$. For the next highlight no gain is detected at $m=1$ and 3, and the ranking is
worse at $m=5$. For comparison, widening the baseline's own scale from 20 to 40, with no similarity
at all, gains \RtApProxFortyVsProxTwentyOneFull{} AP at $m=1$.

\paragraph{Two counterexamples.} Would a gain, even a larger one, show that a method has found what
the reader likes? Not by itself. We ran the same code on synthetic pages whose readers we control
(Table~\ref{tab:counter}; Appendix~\ref{app:data}).

\begin{table}[t]
\centering\footnotesize\setlength{\tabcolsep}{3pt}
\begin{tabular}{@{}llll@{}}
\toprule
 & & & PROXBLEND(40) \\
synthetic pages & KNN $-$ POP & UBCF $-$ POP & $-$ PROXPOP(40) \\
\midrule
position only & \RaNarrowApKnnOneFull & \RaNarrowApUbcfOneFull & \RtsNarrowApProxBlendFortyVsProxFortyOneFull \\
types only & \RsTasteModApKnnOneFull & \RsTasteModApUbcfOneFull & \RtsTasteModApProxBlendFortyVsProxFortyOneFull \\
types and position & \RsPosTasteApKnnOneFull & \RsPosTasteApUbcfOneFull & \RtsPosTasteApProxBlendFortyVsProxFortyOneFull \\
\bottomrule
\end{tabular}
\caption{Three kinds of synthetic page: paired differences in AP at $m=1$ (300 pages of 100
sentences per kind, with the same code and intervals as the data). \emph{Position only}: all readers share
one set of preferences; each reads one window of 15 to 30 sentences and marks inside it in reading
order. \emph{Types only}: two reader types; half of each reader's marking probability is their
type's own; marks fall anywhere on the page, in random order. \emph{Types and position}: the same
types, inside reading windows of 30 to 60 sentences. At $m=3$ and 5 every sign is the same, except
that the similarity scorers' losses on the third kind are not resolved at $m=5$.}
\label{tab:counter}
\end{table}

\emph{A gain without distinct preferences.} On the first kind of page every reader has the same
preferences as everyone else, reads one window of the page and marks inside it in reading order.
Nothing distinguishes these readers but where they read. Both similarity scorers beat popularity
there, and blending similarity into the position baseline raises AP: on such pages similarity
measures position.

\emph{Distinct preferences without a gain.} On the other two kinds there are two reader types, and
half of each reader's marking probability comes from their type's own preferences. Both similarity
scorers lose to popularity on both, and where the types read in windows the blend lowers AP.

A gain of a similarity method, over popularity or over a position baseline, therefore does not by
itself establish that the method has identified reader-specific preferences, and the absence of a
gain does not establish that readers have none. These are counterexamples and not a model of our
pages: they show which signs can occur, not how large an effect to expect, and the outcome depends
on how wide the reading windows are, on the order of marking and on how the types' preferences are
spread.

\section{Out of time order, a method is shown where the reader went}\label{sec:protocol}

The given-$m$ protocol as usually run ignores order: it reveals $m$ of a reader's highlights at
random and asks for the rest. The highlights it reveals then lie among the ones that remain to be
predicted, and not before them, so the reveal hands a method the reader's later positions. To
isolate the protocol we changed it on the time-ordered data, on the same readers (the registered
units whose marks form at least $m+1$ runs of consecutive sentences), so that every difference is
paired.

Revealing $m$ random highlights instead of the first $m$, with the same pool, raises UBCF $-$ POP on
AP by \RvHlMinusTlUbcfThreeFull{} and \RvHlMinusTlUbcfFiveFull{} at $m=3$ and 5, and KNN $-$ POP by
\RvHlMinusTlKnnFiveFull{} at $m=5$ (its rise at $m=3$ is not resolved); at $m=1$ neither is
resolved. A fully order-free protocol, which reveals $m$ random runs of marked sentences and takes
every other reader as the pool, whenever they came, turns UBCF's resolved loss in time order on
these units into a resolved gain at $m=3$ and 5 (at $m=5$, from \AbTlUbcfFiveFull{} to
\AbRaUbcfFiveFull{}).

A random reveal also changes which highlights remain to be predicted. So we held the labels fixed:
keeping the cut, the candidates, the positives and the pool of the registered unit at $m=5$
(\RwFixN{} units), we varied only which one of the newcomer's first five highlights the scorer sees.
Shown the last one instead of the first, KNN gains \RwFixOneKnnApLastVsFirstFull{} AP and UBCF
\RwFixOneUbcfApLastVsFirstFull{}. PROXPOP(20), which takes nothing from the shown highlight but
where it is, gains \RwFixOneProxApLastVsFirstFull{}. Position is therefore part of what a later
highlight gives a scorer; how much of the similarity scorers' gain is positional this contrast
cannot say.

\section{Discussion}\label{sec:discussion}

\paragraph{What to do differently.} The claim of this paper is a claim about evaluation, and four
rules follow. \emph{Name the target first}: after one highlight
the same position rule was far ahead for the next highlight and behind over all later ones.
\emph{Put position in the comparison}, as a rule that ignores other readers and as one that combines
position with popularity; at the two wider scales we tried, neither similarity method was ahead of
the second on either target. \emph{Reveal in time order}: with five highlights revealed, a random
reveal raised both similarity methods on the same readers. \emph{Do not take a gain, by itself, as a
sign of preference}: differences in where readers read can produce one.

\paragraph{For products.} These are notes from an offline evaluation, not tested recommendations.
About half the time, the position rule's top five hold a sentence of the reader's next highlight. We
did not test whether suggesting those sentences is of any use, so we offer the rule as a benchmark
for any predictor of the next highlight, not as a feature.

\paragraph{Limits.} The claim is about particular rules and two targets, not about readers: nothing
here tests whether readers have distinct preferences, or whether the sub-groups an earlier study
described among a document's readers exist~\cite{factions}. We tried neighbourhood methods; a
learned model, or one that reads the text, may beat a position baseline, and beating it would still
not by itself show that the model has found the reader's preferences. The position results are
exploratory, rest on many unadjusted intervals, and are unconfirmed: a confirmation of the
next-highlight result on highlights recorded from 2026-10-07 on, with one test, is specified in the
ancillary files under timestamps from two independent authorities, and has not been run. The data
come from one platform and are anchored to each page's text as fetched now, not as each reader saw
it, which favours unchanged pages. A reader's position is known only through their latest
highlight, and the time between highlights was
not kept, so a next highlight can come seconds or months after the cut. Readers can see earlier
readers' highlights, and when people see what others did, later behaviour follows earlier
behaviour~\cite{muchnik2013}, so some of popularity's strength may be such influence.

\section{Conclusion}

Inside a document, a reader's first highlights say where the reader is, and an evaluation of
personalisation that leaves this out compares with too weak a baseline. With position in the
comparison, two standard similarity methods are behind it on both targets, and a gain over it would
not by itself show that a method has found a reader's preferences. Whether readers resemble
particular earlier readers is not decided by any of this. The position results are exploratory and
await confirmation on new records.

\section*{Ethics and data}
This is a retrospective, non-interventional secondary analysis of service records collected under the
platform's Terms of Service and Privacy Policy, which name research among the purposes of processing.
The study read production records read-only, kept only highlight documents public under the
platform's own rule (public is the default), and wrote pseudonymised records without account
identifiers, highlight text or absolute times. For the check by hand, one author viewed the text and
recorded times of two highlights for each of \InUnits{} sampled units in a local file, which was
then deleted. This is pseudonymised behavioural data, not anonymous data; record-level files are not
released, and every reported result is an aggregate over documents with at least six readers. The
operating company is not a research institution and has no institutional review board; no
institutional review board reviewed this study, and no formal ethics approval was sought. Both
authors are affiliated with the platform's operator.

\section*{AI assistance}
The study design, the extraction and analysis code, the statistics, the figure and a first draft of
this paper were produced with an AI coding agent (Anthropic's Claude) under the authors' direction.
Two AI auditors, OpenAI's Codex and a separate Claude instance, reviewed the design, the code and
the write-ups, and built some of the synthetic pages. The authors take full responsibility for the
content.

\section*{Data and code availability}
Record-level data cannot be released. The ancillary files contain the specifications, the results
write-ups, the analysis code as specification, and every aggregate artifact from which every number
in this paper is generated. They also hold the analyses this paper only mentions or leaves out
(among them an earlier order-free study and further position and blend scorers), the protocol and
the counts of the check by hand, and the specification of the confirmation. The synthetic-page
analyses run from the ancillary files alone.

\appendix

\section{Details of the data, the scoring and the synthetic pages}\label{app:data}

\paragraph{Pages and highlights.} Pages were found by scanning the platform's stored URLs with at
least three nominal readers and grouping URL variants under one canonical page key. A reader counts
towards a page's six when they have a \emph{public} highlight document whose own URL is the page.
Imported and accepted-suggestion highlights, image highlights, spans under twelve characters and
repeated spans are dropped before ordering, so a reader's ``first'' highlights are the first that
survive these filters.

\paragraph{Text and anchoring.} A page's text is fetched now by a process without credentials (a
cached copy only where the fetch fails or gives under 30 sentences) and split into sentences with
one fixed segmenter. Of the \TmGroupsAll{} eligible page groups, \TmGroupsNoText{}
(\TmNoTextShare{}) had no usable text and \TmGroupsIdentical{} were dropped because their text was
identical to another group's. The other \TmGroups{} hold \TmHighlights{} highlights, of which
\TmAnchored{} (\TmAnchoredShare{}) anchor to a sentence of today's text. The pages we analyse are
the \TmAnchoredPages{} on which at least six readers have an anchored highlight. Of the
\RtAnchoredTotal{} anchored highlights on them, \RtAnchoredOneShare{} anchor to one sentence,
\RtAnchoredRunShare{} to a contiguous run of several, and \RtAnchoredScatterShare{} to sentences
that are not contiguous (the text matches in more than one place, or spans a very short sentence,
which is never matched). Of the matches between a highlight and a
sentence, \RxFuzzyShare{} were made by token overlap alone.

\paragraph{Units.} At $m=1$, of the \RwFunReadersOne{} reader-page pairs on the analysed pages,
\RwFunShortOne{} made a single highlight, \RwFunUnanchOne{} had a first highlight that did not
anchor, \RwFunNoPosOne{} added no sentence afterwards and \RwFunSmallPoolOne{} had fewer than five
earlier readers. That leaves the \TmUnitsOne{} units on \TmPages{} pages.

\paragraph{Ties.} KNN breaks ties among equally similar readers by a seeded random order. Ties in a
scorer's own key are broken by pool popularity (AHEAD uses none; the blend's are broken by its
position baseline first), and what is still tied is scored in expectation over a random order.

\paragraph{The check by hand.} The \InUnits{} units are \InPerStratum{} from each of three strata,
by how the newcomer's first highlight anchored (to one sentence by containment; to several
sentences, at least one by containment; by token overlap only), drawn by a keyed hash from a fresh
read of the same pages' highlights. The protocol and the counts are in the ancillary files.

\paragraph{Synthetic pages.} Each kind has 300 pages of 100 sentences with 8 to 40 readers, one
sentence per highlight, and a page-wide probability of being marked that is drawn for each sentence
from a Beta(0.6, 3) distribution. A reader with a reading window marks only inside it, in reading
order, each sentence with twice its probability (capped at one). A reader of a type marks a sentence
with half the page-wide probability plus half their type's own, which is drawn from the same
distribution. The kinds of Table~\ref{tab:counter} are three of ten we ran, which vary the width of
the windows, the order of marking and the strength and spread of the types; the ancillary files
report all ten, and call these three ``position only, narrow windows'', ``taste, moderate'' and
``position and taste''. Across them, PROXBLEND(40) $-$ PROXPOP(40) on AP takes each sign both with and
without reader types.

\end{document}